\documentclass[twocolumn,aps,nofootinbib]{revtex4-1}
\usepackage{graphicx}
\usepackage{booktabs}
\usepackage{xcolor}
\usepackage{hyperref}
\usepackage{multirow}
\usepackage{setspace}
\usepackage{amsmath, amssymb, amsthm}
\usepackage{cleveref}
\usepackage{physics}
\usepackage{url}
\newcommand{\comment}[1]{}

\begin{document}
\author{Alexander F. Siegenfeld$^{1,2}$} 
\email{Corresponding author.  Email: asiegenf@mit.edu}
\author{Yaneer Bar-Yam$^2$}
\affiliation{$^1$Department of Physics, Massachusetts Institute of Technology,  Cambridge, MA}
\affiliation{$^2$New England Complex Systems Institute, Cambridge, MA}
\title{Eliminating COVID-19: The Impact of Travel and Timing}

\begin{abstract}
We analyze the spread of COVID-19 by considering the transmission of the disease among individuals both within and between regions.   A set of regions can be defined as any partition of a population such that travel/social contact within each region far exceeds that between them.  COVID-19 can be eliminated if the \textit{region-to-region reproductive number}---i.e. the average number of other regions to which a single infected region will transmit the virus---is reduced to less than one.  We find that this region-to-region reproductive number is proportional to the travel rate between regions and exponential in the length of the time-delay before region-level control measures are imposed.  Thus, reductions in travel and the speed with which regions take action play decisive roles in whether COVID-19 is eliminated from a collection of regions.  If, on average, infected regions (including those that become re-infected in the future) impose social distancing measures shortly after active spreading begins within them, the number of infected regions, and thus the number of regions in which such measures are required, will exponentially decrease over time.  Elimination will in this case be a stable fixed point even after the social distancing measures have been lifted from most of the regions.
\end{abstract}

\maketitle

\textbf{The outbreak of COVID-19, caused by the novel coronavirus SARS-CoV-2, emerged in Wuhan, China in December 2019~\cite{WHO1} and has since become a severe pandemic~\cite{WHO2}.  Understanding the dynamics of disease transmission both within and between regions can provide insight into how to eliminate the outbreak~\cite{Watts11157,rauch2006long,colizza2007reaction,ajelli2010comparing,tanaka2014random,ball2015seven}.  
While the assumptions of all models fail to describe the details of complex real-world systems, these systems may possess large-scale behaviors that do not depend on all these details~\cite{Kardar2007,bar2016big,alex2019introduction}.    
For COVID-19, the parameters for these large-scale behaviors depend on the scale of the description (e.g. the size of the regions) and include the average size of an outbreak within a region and the transmissibility of the outbreak between regions.  The values of these two parameters, both of which can be controlled with interventions, determine whether the multi-region behavior of the epidemic is that of exponential spread until saturation (e.g. herd immunity) or exponential decay until elimination.  We find that reductions in travel and the speed with which regions take action play decisive roles in whether COVID-19 is eliminated from a collection of regions.  If, on average, infected regions (including those that become re-infected in the future) impose control measures shortly after active spreading begins within them, the number of infected regions, and thus the number of regions in which such measures are required, will exponentially decrease over time.}

A central concept in the study of disease spread is the reproductive number, i.e. the average number of people to whom an infected individual will transmit the disease~\cite{diekmann1990definition}.  Outbreaks can be stopped if interventions reduce this reproductive number to less than one.  Here, we consider the analogous \textit{region-to-region reproductive number} $R_*$~\cite{ball1997,colizza}, defined as the average number of other regions (including those that have been previously infected) to which a single infected region will transmit the infection.  
Our analysis applies to a set of regions defined as any partition of a population such that travel/social contact within each region far exceeds that between them (e.g. the U.S. could be partitioned by state or commuting zone boundaries).  For the purposes of interventions, treating larger areas as single regions means that social distancing measures will be homogeneously applied to larger areas but also means that it is easier to achieve lower travel rates between such areas.  The coronavirus can be eliminated if regions that become infected contain the coronavirus rapidly enough for $R_*<1$ to be achieved.  

\begin{figure}
\centering
\includegraphics[width=.45\textwidth]{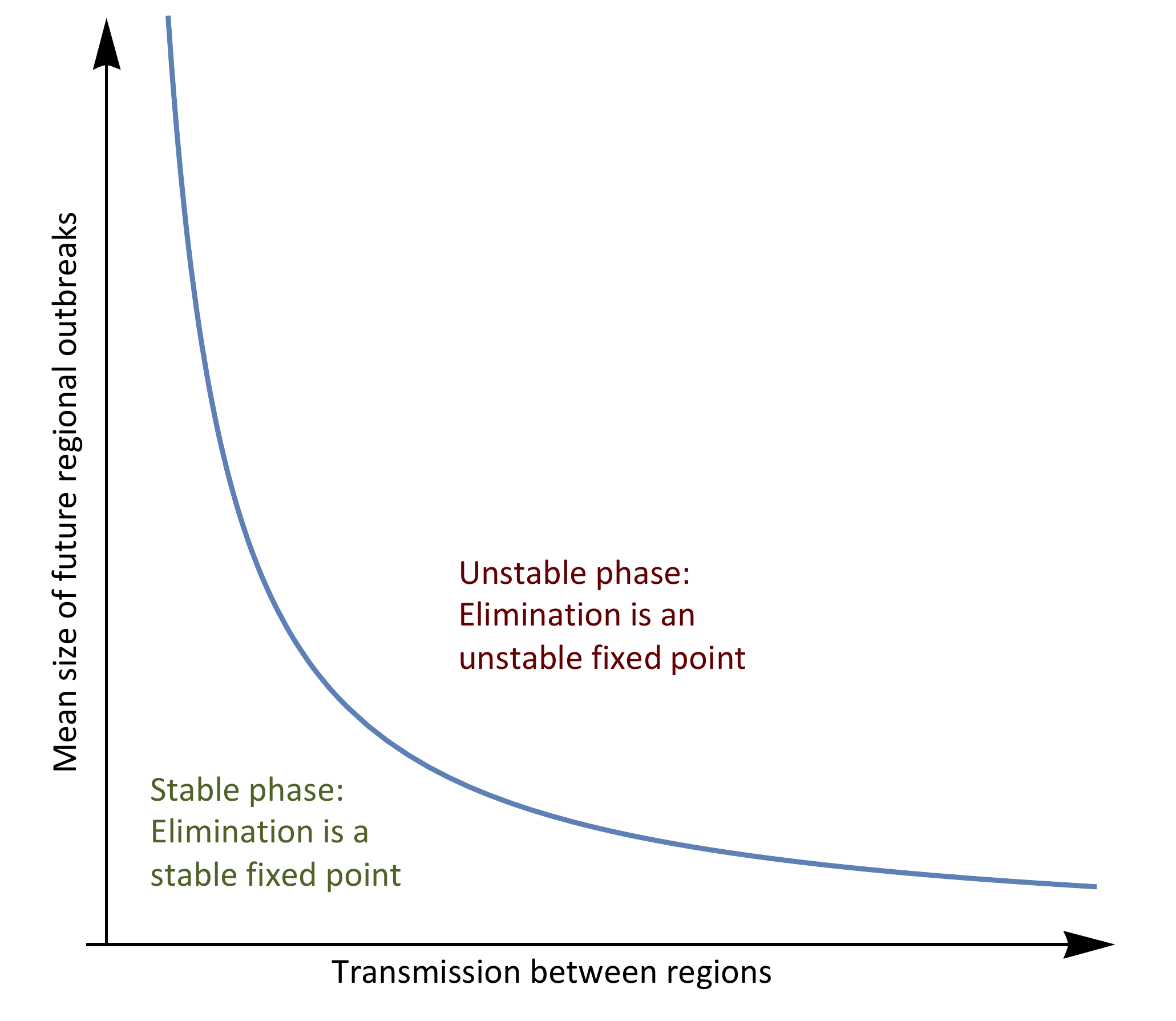}
\caption{A collection of geographic regions can exist in one of two regimes with respect to COVID-19.  If the virus is eliminated from currently infected regions, the question then becomes whether or not this elimination is stable.  The stability of elimination depends on (1) the average total number of cases that will result from the disease being transmitted to a region, which in turn depends on (among other factors) how quickly regions locally lock down if they are infected or re-infected, and (2) the probability that an infected individual in one region will infect an individual in another, which in turn depends on the rate of travel between regions.}
\label{fig:phases}
\end{figure}

The disease is modeled as being transmitted among individuals within a region, with travel allowing the disease to spread between regions.  We define a region as \textit{infected} if someone with the infection enters the region.  Conditioning on region $c$ being infected, we let $N_c$ be a random variable (that could be zero) denoting the total number of infections that occur in region $c$ from the time of infection to the time at which the virus is eliminated or contained.  Let $p_c$ be the probability that an infected individual in region $c$ will travel to another region during the period in which that individual is contagious.  Then, the region-to-region reproductive number for region $c$---which we define as the expectation of the number of regions that region $c$ will infect if it becomes infected---is $R_*^c=\mathbb{E}[N_c]p_c$.\footnote{We note that if region $c$ were infected multiple times, $\mathbb{E}[N_c]$ would be higher than if it were infected once, but it is assumed that infecting an already infected region will on average contribute no more to disease spread than infecting a currently uninfected region. Thus, like the basic reproductive number, this region-to-region reproductive number overestimates the disease spread away from the limit of most regions being uninfected by counting a region that has been infected multiple times during a single outbreak as multiple regions being infected.  We also note that, as in SIS compartmental models, an infected region for which the virus is contained can later be re-infected.} 
$R_*^c$ may differ from region to region, but if the interventions are sufficiently fast and strong such that $R_*$, the average value of $R_*^c$ over a set of regions with each region weighted by its probability of being infected~\cite{diekmann1990definition}, is less than one, then the outbreak will not be self-sustaining within that set of regions.  Put another way, a set of regions can exist in one of two regimes: a regime for which elimination is a stable fixed point of the system and a regime for which it is unstable (see \cref{fig:phases}).   
A change in policy can shift a collection of regions from the unstable regime ($R_*>1$) to the stable regime ($R_*<1$) or vice versa.  Although the values of $N_c$ for a set of regions could currently be high, the current value of $R_*$ is determined by $\mathbb{E}[N_c]$ and $p_c$ for the regions that will be infected or re-infected in the future after social distancing has eliminated the virus from currently infected regions.

We now describe a simple model to estimate $N_c$, not to provide a precise description of the epidemic trajectory but rather to clarify how various interventions may affect outbreak size.  The one key assumption is that the reproductive number within individual regions can be reduced below one with sufficiently strong social distancing measures, as suggested by empirical evidence (see below).

Let $i_0^c$ be a stochastic factor that roughly corresponds to the initial foothold that the virus gains in region $c$ conditioning on an infected individual entering the region, with $i_0^c=0$ corresponding to the case in which no one was infected or a few people were infected but the outbreak was contained by contact tracing/quarantine or otherwise spread no further.
If the outbreak is contained ($i_0^c=0$), then the number of active infections remains at zero for the purposes of this model because---by definition---the outbreak has no chance of spreading to other regions. 
If the outbreak is not contained ($i_0^c\neq 0$), the number of active infections is modeled as growing with time $t$ at an exponential rate $e^{r_ct}$.  After time $T_c$ (the delay in response), the region implements aggressive social distancing measures that cause the number of active infections to decay as $e^{-t/\tau_c}$.\footnote
{The assumption of exponential increase followed by exponential decay after intervention assumes that the proportion of susceptible individuals is roughly constant, i.e. that the region intervenes before a significant fraction of its population is infected, which is the regime with which we are concerned.  (To the extent that this assumption does not hold, its use will result in an overestimate of the number of infected individuals and thus does not affect our main conclusions.)}
Such exponential decrease will occur if the aggressive social distancing measures, together with testing, contact tracing, and quarantine, can reduce the reproductive number ($R$) of the virus below one.  
The greater the reduction in $R$, the smaller the value of $\tau_c$ and thus the faster the decrease in infections.\footnote
{For $R<1$, $R$ is related to $\tau_c$ by $1=\int_0^\infty Rg(t)e^{t/\tau_c}dt$ where $g(t)$ is the distribution of generation intervals~\cite{wallinga2007generation}.}

\begin{figure}
\centering
\vspace{2em}
\includegraphics[width=.45\textwidth]{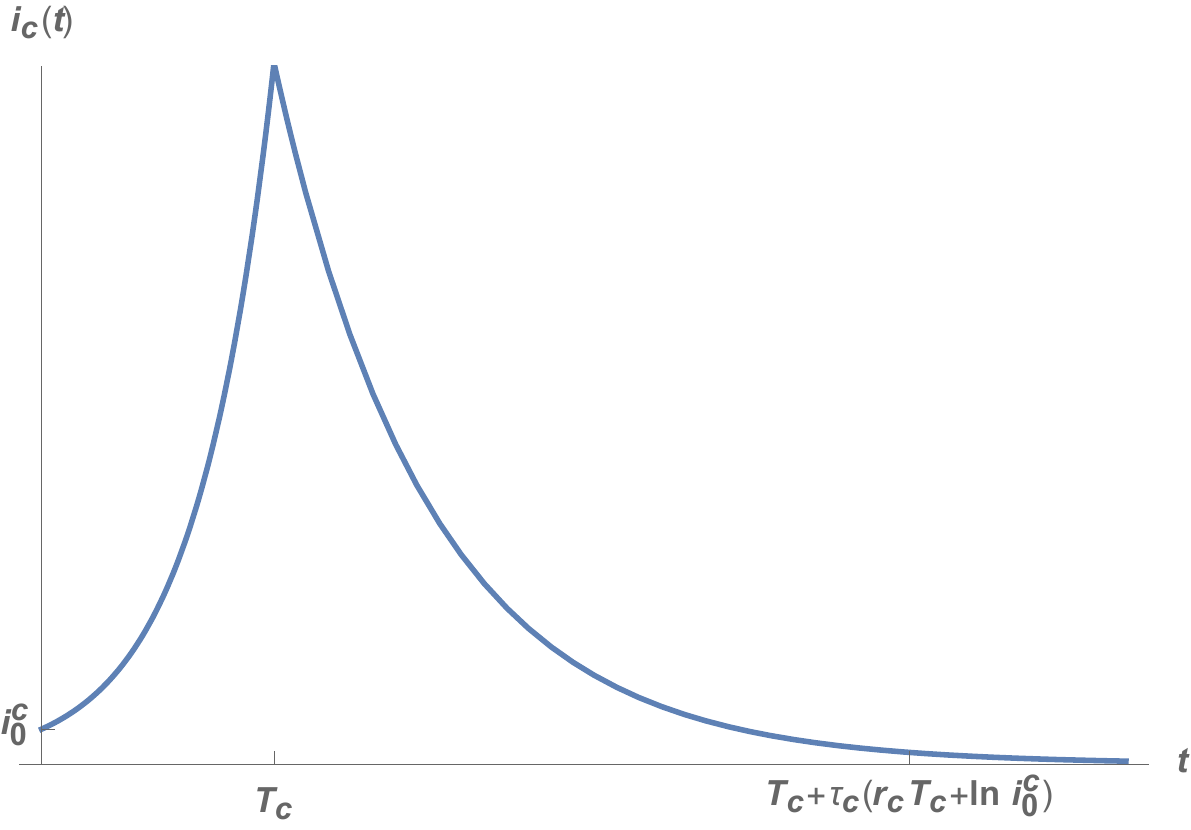}
\caption{The number of active infections in a region as a function of time.  After time $T_c$, aggressive social distancing measures are implemented.  They must remain in place for a duration greater than $\tau_c(r_cT_c+\ln i_0^c)$ (see \cref{eq:dur}).  Thus, the longer the region waits to enact the measures, the longer the total amount of time they must remain in place.}
\label{fig:i}
\end{figure}

The number of active infections in region $c$ as a function of time (see \cref{fig:i}) can therefore be written as:
\begin{equation}
i_c(t)= \begin{cases} 
      i_0^ce^{r_ct} & t\leq T_c \\
      i_0^ce^{r_cT_c}e^{-(t-T_c)/\tau_c} & t\geq T_c
   \end{cases}
   \end{equation}

The social distancing measures can be lifted once no active infections remain in the region or once all active infections have been contained.  Solving for $i_c(t)=1$ (assuming $i_0^c\neq0$) yields an approximate duration for the social distancing measures of
\begin{equation}
\label{eq:dur}
\tau_c(r_cT_c+\ln i_0^c)
\end{equation}
As the number of cases becomes increasingly small, contact tracing may become increasingly effective and hasten the drop of $i_c(t)$ to zero.  The probability that the number of infections will rebound after the social distancing measures are lifted---in which case an additional phase of such measures will be needed, as in the Imperial College report~\cite{fergusonimpact}---will depend on the probability of importation from other regions, which in turn will depend on the region-to-region reproductive number $R_*$.  Even though the virus can be re-imported, as long as $R_*<1$ the number of infected regions will on average decrease over time, since re-importation events are included in $R_*$.

Each infected region $c$ infects a currently uninfected region with a probability rate proportional to the number of active infections $i_c(t)$ times the probability rate $p_c$ that an infected individual will travel to an uninfected region.
The number of new infected regions spawned by region $c$ can thus be modeled as a Poisson process with rate $i_c(t)p_c$.
This modeling assumption overestimates the spread of the disease to new regions by counting a single new region that has been infected multiple times as multiple new infected regions.  Note that a single new region that is infected by more than one other region is also counted as multiple new infected regions.  The smaller the number of regions infected and the smaller the probability of one region infecting another, the smaller the probability that the same region will be infected twice.  Nonetheless, for certain network connectivities, this model may overestimate $R_*$.  (Our main conclusions are unaffected because $R_*$ will be less than one if its overestimate is.)   

Let $p_0^c$ be the per capita probability rate before time $T_c$ of individuals in region $c$ traveling to other regions and $p_1^c$ be the probability rate afterwards ($p_1^c$ will be less than $p_0^c$ if travel is discouraged and/or restricted at the time social distancing measures are implemented).   The number of new regions that are infected by region $c$ will then be a Poisson random variable with a mean of 
\begin{align}
i_0^c p_0^c&\int_0^{T_c} e^{r_c t}dt+i_0^c p_1^ce^{r_cT_c} \int_0^{\tau_c(r_cT_c+\ln i_0^c)} e^{-t/\tau_c}dt \\
&=\begin{cases}
i_0^c\qty(p_0^c\frac{e^{r_cT_c}-1}{r_c}+p_1^c\tau_c(e^{r_cT_c}-\frac{1}{i_0^c})) & i_0^c>0 \\
0 & i_0^c=0
\end{cases}
\end{align}
Taking the expected value over $i_0^c$ (and allowing for a slight overestimate of $R_*^c$ by treating the $\frac{1}{i_0^c}$ term as negligible when $i_0^c>0$) yields
\begin{equation}
\label{eq:Rc}
R_*^c=\mathbb{E}[N_c]p_c=\mathbb{E}[i_0^c]\qty(p_0^c\frac{e^{r_cT_c}-1}{r}+p_1^c\tau_c e^{r_cT_c})
\end{equation}

\begin{figure}
\centering
\vspace{2em}
\includegraphics[width=.45\textwidth]{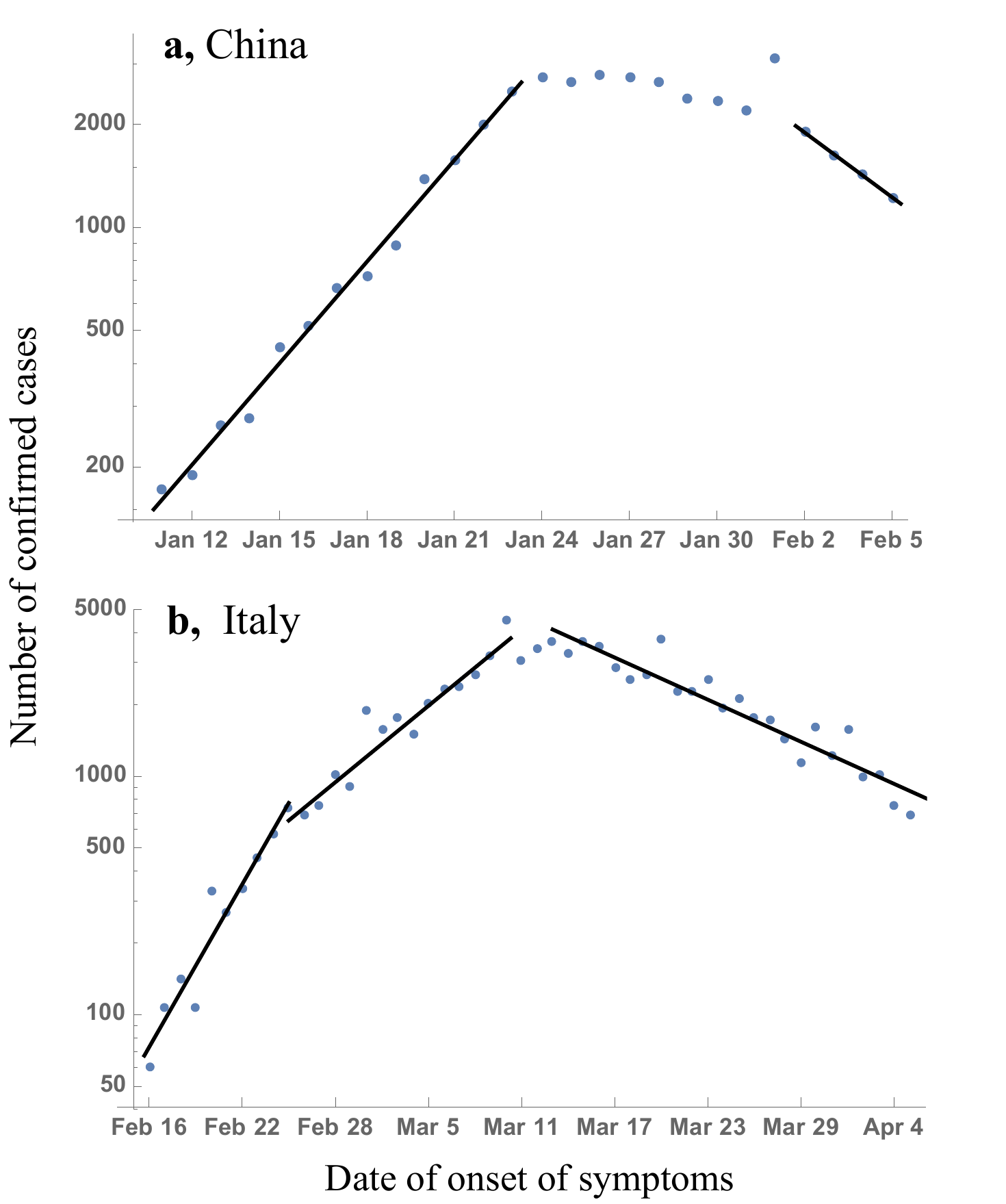}
\caption{\textbf{a,} Log plot of the daily number of confirmed cases in China by date that these patients self-reported as the onset of their symptoms.  The best OLS linear fits to the natural log of the number of cases are shown: For Jan. 11-23 (up until the lockdown), the slope (in units of day$^{-1}$) is 0.228 ($R^2=0.991$, 95\% confidence interval (CI) $[0.214,~0.242]$), which corresponds to a doubling time of 3.04 days.  For Feb. 2-5 the slope is $-0.145$ ($R^2=0.999$, 95\% CI $[-0.160,~-0.131]$), which corresponds to a halving time of 4.78 days.  
Data are from ref.~\cite{jama}, which includes cases diagnosed through Feb. 11.  Not pictured: There is a drop in cases with onsets of symptoms after Feb. 5, likely due to many of those cases being diagnosed after Feb. 11.  It should be noted that the number of cases by date of diagnosis (not pictured) continues to increase through Feb. 4, indicating that in general the date of diagnosis substantially lags the date of onset of symptoms, which itself lags the date of infection.  Thus, a considerable amount of time can pass before even an immediate decrease in the infection rate can be observed.\newline
\textbf{b,} Log plot of the daily number of confirmed cases in Italy by date of symptom onset (data are from ref.~\cite{italy}).  The best OLS linear fits are shown and have slopes (in units of day$^{-1}$) of 0.262 ($R^2=0.927$, 95\% CI $[0.202,~0.322]$) for Feb. 16-25, 0.123 ($R^2=0.923$, 95\% CI $[0.102,~0.144]$) for Feb. 25 - Mar. 10, and $-0.068$ ($R^2=0.901$, 95\% CI $[-0.078,~-0.058]$) for Mar. 13 - Apr. 5.  The change in the exponential growth rate from 0.262 to 0.123 likely occurred due to partial measures implemented by Italy, but it was not until a nationwide lockdown was implemented on March 9 that exponential growth changed to exponential decline.  The rate of decline is much slower in Italy than in China, perhaps due to China's stronger lockdown enforcement and contact tracing/quarantine measures.}
\label{fig:onset}
\end{figure}

\begin{figure}
\centering
\vspace{2em}
\includegraphics[width=.5\textwidth]{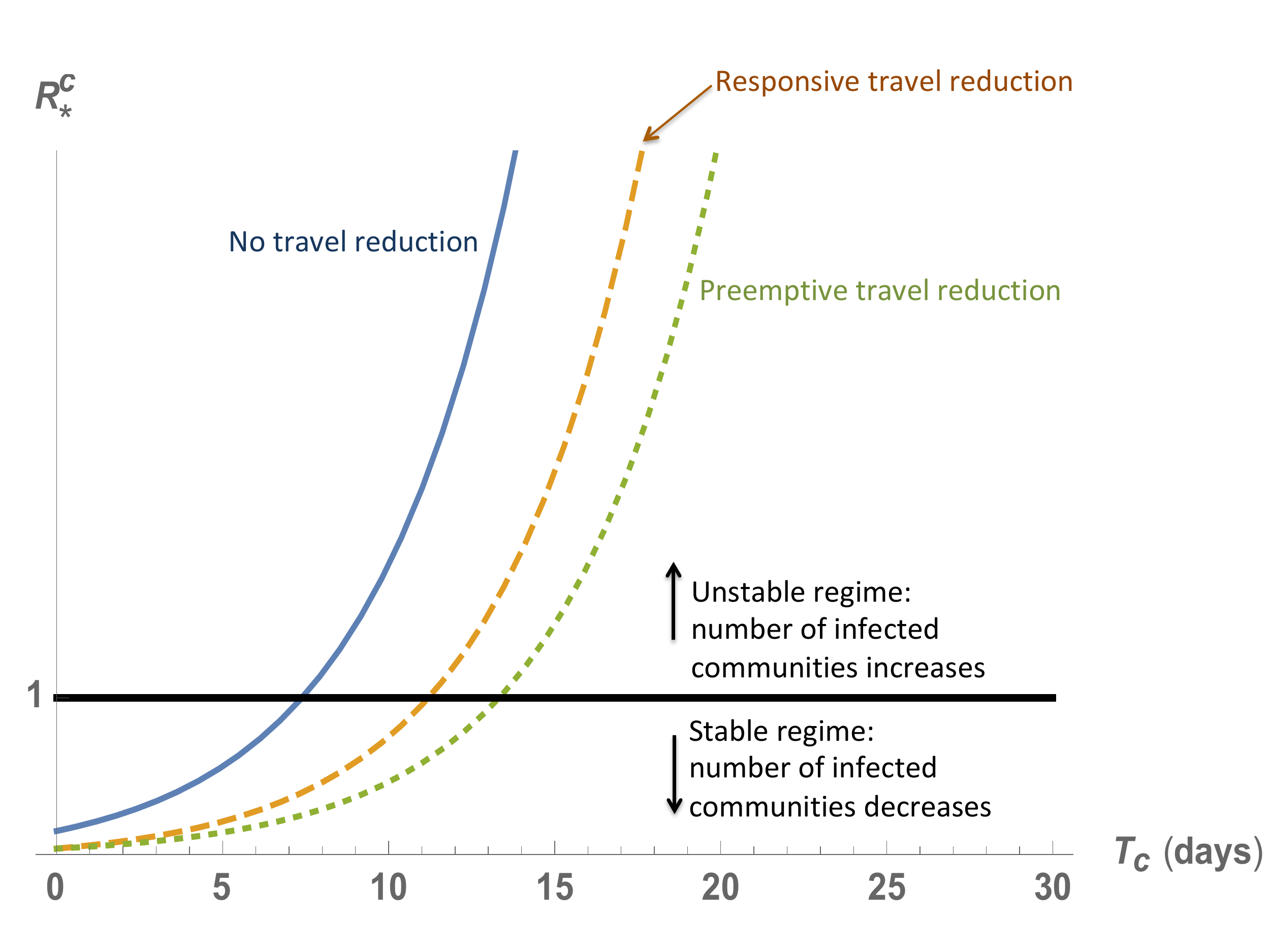}
\caption{Dependence of $R_*^c$ (the average number of regions to which region $c$ will transmit the disease) on $T_c$ (the time-delay before the social distancing measures are enacted).  If $R_*$ (a weighted average of $R_*^c$) is less than one, the number of infected regions will exponentially decrease and the disease will be eliminated (and the smaller $R_*$ is, the faster it will be eliminated); otherwise, the number of infected regions will increase until saturation.  \newline \textbf{Parameter values:} All curves use $\mathbb{E}[i_0^c]=2.5$, $\tau_c=15$ days, and $r_c=.228$ day$^{-1}$.  Solid curve: no travel reduction, $p_0^c=p_1^c=0.004$ day$^{-1}$.  Dashed curve: 4-fold (responsive) travel reduction after time $T_c$, $p^c_0=0.004$ day$^{-1}$ and $p_1^c=0.001$ day$^{-1}$.  Dotted curve: general (preemptive) 4-fold travel reduction: $p_0^c=p_1^c=0.001$ day$^{-1}$.}
\label{fig:R}
\end{figure}

In order to better understand the extent of the measures required to achieve $R_*<1$, we estimate the values of the parameters in \cref{eq:Rc} (see \cref{fig:onset} and Methods) in order to determine $R_*^c$ as a function of the time-delay before aggressive social distancing measures are enacted, as shown in \cref{fig:R}.  Note that the time-delay is measured from the time at which exponential growth begins to occur---which could be as early as the first infection transmitted within the region---not the time at which exponential growth is first measured.

Reducing $R_*$ to less than one will stop the outbreak, and if $R_*$ is already less than one, further reducing $R_*$ will increase the speed at which the outbreak is stopped.  Since $R_*$ is proportional to both $\mathbb{E}[N_c]$ and the travel rates between regions, any intervention that reduces the sizes of outbreaks within regions and/or travel between them will also reduce $R_*$.  In the language of \cref{eq:Rc}: 
\begin{itemize}
\item A reduction in travel from region $c$ results in a linear reduction in $R_*^c$ through $p_0^c$ and $p_1^c$.  
\item Improvements in testing, contact tracing, and quarantine reduce $R_*^c$ through $\mathbb{E}[i_0^c]$, $r_c$, and $\tau_c$.
\item The preemptive reduction of large events such as conferences (as well as general social distancing) reduces the probability of a super-spreader event as well as general transmission, reducing both $\mathbb{E}[i_0^c]$ and $r_c$.
\item Reductions in $r_cT_c$ not only exponentially reduce $R_*^c$ (as well as the total number of infections within the region) but also linearly reduce the amount of time for which the social distancing measures must remain in place.
\item Augmenting social distancing measures (after time $T_c$) decreases $\tau_c$, which results in a linear decrease in both $R_*^c$ and the time for which the distancing measures must remain in place.   
\end{itemize}

We conclude with a few comments.  First, without the timely implementation of aggressive social distancing measures, restricting travel from infected regions serves only to delay the spread of the outbreak, as found in other studies~\cite{bajardi2011,poletto2014,brownstein2006,germann2006,epstein2007,wells2020,chinazzi2020}.  However, when a reduction in travel is coupled with social distancing measures, the travel reduction will not only delay the spread of the outbreak but in some cases will also be the determining factor in whether or not the outbreak is eliminated.  (If $R_*<1$ can be achieved without reducing travel, travel reductions can greatly decrease the duration and total case count of the outbreak by further reducing $R_*$.)  Empirically, travel restrictions, when combined with other sufficiently strong interventions, have been predicted~\cite{chinazzi2020} and found~\cite{kraemer2020} to substantially curb the COVID-19 epidemic.  
  
Second, because $R_*^c$ depends exponentially on $T_c$, the longer a region waits to take aggressive social distancing measures, the more important it becomes to act without delay.  It is important to note, however, that there is no advantage to delaying at all.  Immediately implementing aggressive social distancing measures as soon as there is evidence of the disease spreading within the region will not only reduce the total amount of time for which such measures must remain in place but will also exponentially reduce the probability of infecting or re-infecting another region.  Practically speaking, all regions within which the virus is spreading must implement and maintain control measures until the virus has been eliminated or contained.  If the virus is then re-imported into a locale, sufficiently strong interventions should be implemented in a region around that locale that is large enough such that the per capital rate of travel out of that region is sufficiently low (e.g. predominantly by flight).  Practically, this may involve locking down regions the size of commuting zones (which may span multiple states), and possibly entire states.  Per capita travel between regions does not have to be zero, but it must be small (the precise value depends on the parameters in \cref{eq:Rc}; \cref{fig:R} gives a rough estimate of what is necessary). 

Finally, just as transmission within a region is an exponential process for which the need for immediate action is not always apparent, the transmission between regions is also such an exponential process.  At first the number of infected regions is deceptively small, but without rapid action this number exponentially grows.  The sooner a set of regions decides to adopt a protocol that reduces $R_*$ below one, the shorter the amount of time between the adoption of the protocol and the elimination of the disease.

\begin{acknowledgments}
This material is based upon work supported by the National Science Foundation Graduate Research Fellowship Program under Grant No. 1122374 and by the Hertz Foundation.  We thank Maxim Rabinovich for helpful early discussions about the model and Edward Kaplan, Philip Welkhoff, Chen Shen, and Daniel Klein for helpful comments.
\end{acknowledgments}

\section*{Methods}
In this section we estimate the parameters in \cref{eq:Rc} for COVID-19.

The doubling time of the epidemic can vary from location to location and depends on pre-lockdown interventions (see e.g. the change in the growth rate for Italy in \cref{fig:onset}), but using the number of confirmed cases in China by date of symptom onset (rather than by date of diagnosis)~\cite{jama} yields a doubling time of 3.04 days in the period leading up to the Jan. 23 lockdown, which corresponds to $r=0.228~\text{day}^{-1}$ (\cref{fig:onset}).\footnote
{Some studies estimated the doubling time for COVID-19 at approximately 7 days~\cite{wu2020nowcasting,li2020early}, but even a 5 day doubling period is implausibly long, given that in various countries, even with some preventative measures, the number of infections has increased by far more than a factor of 64 over 30 days~\cite{worldometer}.  Part of the difficulty in estimating the doubling time from the initial period of transmission is that `super-spreader' events may play an important role in the transmission process.  The presence of super-spreader events indicates that the transmission process may be fat-tailed and therefore standard statistical approaches may underestimate the rate of spread when the total number of cases is still small~\cite{Taleb2019}.}  

From this growth rate in China before the Jan. 23 lockdown, the basic reproductive number $R_0$ (which also varies by location) can be calculated from the distribution of generation intervals~\cite{wallinga2007generation}.\footnote
{Empirically, we generally observe the distribution of serial intervals (the times between the onsets of symptoms in two successive cases in a transmission chain) rather than the distribution of generation intervals (the times between two successive infections in a transmission chain).  The means of the two distributions should, however, be the same.} 
Empirical data from various sources support a distribution of generation intervals with a mean of approximately 4.0 to 4.7 days~\cite{NISHIURA2020,du2020serial,zhao2020preliminary,you2020estimation,tindale2020transmission,ganyani2020estimating}, which yield upper bounds of $R_0<e^{4.0r}=2.5$ to $R_0<e^{4.7r}=2.9$.  These upper bounds assume all transmission occurs at the mean generation interval; the spread of generation intervals given a fixed mean results in lower values for $R_0$ given a fixed $r$ (or higher values for $r$ given a fixed $R_0$).  For instance, the serial interval data with a mean of 4.7 days best fit a lognormal distribution (S.D.=2.9 days)~\cite{NISHIURA2020}, which yields $R_0=2.5$.\footnote
{This value was obtained by approximating the generation interval distribution by the serial interval distribution.  The value is consistent with some previously reported $R_0$ values that were based on a mean generation interval overestimated as approximately that of SARS (8.4 days)/MERS (7.6 days)~\cite{liu2020reproductive} because the doubling time was also overestimated by a similar factor.}
This short mean generation time of 4.0 to 4.7 days may explain the difficulty in containing the virus through contact tracing alone.

The values of $\tau_c$ that can be achieved depend on the effectiveness of the social distancing measures.  The data from China (see \cref{fig:onset}) indicate a halving time of as few as 4.78 days is achievable, which corresponds to $\tau=6.9$ days.  However, as a more conservative estimate, we use the data from Italy, which yield $\tau=15$ days.    

$\mathbb{E}[i_0^c]$ is the expected ``effective'' number of people an infected traveler will infect while visiting region $c$, taking into account containment efforts.\footnote
{For instance, if the outbreak is contained such that exponential growth never occurs, the effective number of people infected by the traveler is zero, even if the traveler did infect some individuals in region $c$.}
We estimate $\mathbb{E}[i_0^c]=R_0$; the degree to which $\mathbb{E}[i^0_c]$ differs from $R_0$ depends on how likely a typical traveler is to transmit the virus relative to a typical resident, as well as on the effectiveness of contact tracing and other containment efforts.

The value of $p_c$ depends on the frequency of travel out of region $c$.  As previously noted, there is some choice in how to model the partition of a population into a set of regions.   In general, the larger the regions, the lower the frequency of per capita travel out of them but the more homogeneous the application of the social distancing measures.  Since $p_c$ is smaller for larger regions and $N_c$ is not strongly affected by the size of regions, $R_*$ will be lower if larger regions are chosen, but at the cost of the social distancing measures being applied over larger areas for each new outbreak.
Considering a set of regions within the U.S. that are large enough such that travel between the regions is predominantly by flight yields a per capita travel rate of 0.004 flights out of a region per person per day.\footnote
{This estimate is obtained by dividing the 1.01 billion total passengers traveling by plane to, from, or within the U.S. in 2018~\cite{travel} by the 2018 U.S. population and the number of days in 2018, and then also dividing by 2 so that only flights out of and not into a region are counted.  Using this estimate for $p_0^c$ assumes that the probability that an infected individual will travel equals that of the general public.}

\bibliography{refs}{}
\bibliographystyle{naturemag}

\end{document}